\documentclass{article}
\usepackage{spconf,amsmath,amssymb,graphicx}
\usepackage{pstricks,pstricks-add,pst-node,pst-tree}
\usepackage[bookmarks=true, bookmarksnumbered=true, bookmarksopen=true,
  unicode=true, colorlinks=true,
  pagebackref=true,
  linkcolor=blue,citecolor=blue,filecolor=blue,urlcolor=blue,
  pdfstartview=FitH
]{hyperref}
\usepackage{color}

\title{AUDIOVISUAL SPEAKER DIARIZATION OF TV SERIES}
\name{Xavier Bost, Georges Linar\`es, Serigne Gueye\sthanks{This work
    was partially supported by the French National Research Agency
    (\textsc{anr}) \textsc{contnomina} project (\textsc{anr}-07-240)
    and the Research Federation \textit{Agorantic}, Avignon
    University.}}  \address{LIA, University of Avignon, 339 chemin des
  Meinajari{\`e}s, 84000 Avignon, France}
\begin{document}
\ninept
\maketitle
\begin{abstract}
  Speaker diarization, also known as the ``who spoke when?'' task, may
  be difficult to achieve when applied to narrative films, where
  speakers usually talk in adverse acoustic conditions: background
  music, sound effects, wide variations in intonation may hide the
  inter-speaker variability and make audio-based speaker diarization
  approaches error prone. On the other hand, such fictional movies
  exhibit strong regularities at the image level, particularly within
  dialogue scenes. In this paper, we propose to perform speaker
  diarization within the dialogue scenes of \textsc{tv} series by
  combining the audio and video modalities: speaker diarization is
  first performed by using each of these modalities; the two resulting
  partitions of the instance set are then optimally matched, before
  the remaining instances, corresponding to cases of disagreement
  between the modalities, are finally processed. The results obtained
  by applying such a multi-modal approach to fictional films turn out
  to outperform those obtained by using a single modality.
\end{abstract}
\begin{keywords}
Speaker diarization, multi-modal fusion, video structuration
\end{keywords}

\vspace{2mm}

\noindent \textcolor{red}{\textbf{Cite as:}\\X.~Bost, G.~Linar\`es, S.~Gueye.\\\href{https://ieeexplore.ieee.org/document/7178882}{Audiovisual speaker diarization of TV series.}\\2015 IEEE International Conference on Acoustics, Speech and Signal Processing (ICASSP).\\doi: \href{https://doi.org/10.1109/ICASSP.2015.7178882}{10.1109/ICASSP.2015.7178882}}

\section{INTRODUCTION}
\label{sec:intro}

Speaker diarization (\textsc{sd}) consists in assigning the spoken
segments of an audio stream to their respective speakers, without any
prior knowledge about the speakers involved nor their number. Most of
state of the art systems rely on a two-step approach, performing first
speaker turn detection followed by single-speaker segment clustering.
This last stage is usually based on hierarchical clustering
(\cite{evans2012comparative}) and, more recently, mathematical
programming (\cite{dupuy2012vectors,bredin2013integer}).


\textsc{sd} was first applied to audio-only streams produced in
adverse, but controlled, conditions, such as telephone conversations,
broadcast news, meetings... More recently, \textsc{sd} was extended to
video streams, facing the critical issue of processing contents
produced in an uncontrolled and variable environment.

In~\cite{clement2011speaker}, the authors apply standard speaker
diarization tools to the audio channel of various video documents
collected on the web, resulting in high Diarization Error Rates
(\textsc{der}) in comparison to the scores obtained with the usual
audio streams.

Recent works intend to perform speaker diarization of such audiovisual
contents by using jointly the multiple sources of information conveyed
by multimedia streams. The multi-modal speaker diarization method
introduced in~\cite{Friedland2009} relies on an early combination of
audio and video \textsc{gmm}s, before applying a standard
\textsc{bic}-based agglomerative algorithm to the resulting
features. This technique is evaluated on the~\textsc{ami} corpus which
gathers audiovisual recordings of four speakers playing roles in a
meeting scenario.

In~\cite{bendris2013unsupervised}, the authors make use of both face
clustering and speaker diarization to perform face identification in
\textsc{tv} debates: face clustering and speaker diarization are first
processed independently. Then, the current speaker is identified by
selecting the best modality. Finally, local information about the
current speaker identity are propagated to the whole cluster of the
corresponding utterance.

In~\cite{bredin2013integer}, the authors make use of an intermediate
fusion approach to guide speaker diarization in~\textsc{tv} broadcast
by adding to the set of speech turns new instances originating in
other sources of information: the names written on the screen when a
guest on a reporter is introduced as well as the corresponding
identities. Adding such instances allows to constrain the clustering
process leading to purer classes of speakers.

Finally, audio-based \textsc{sd} has already been applied
(\cite{bredin2012segmentation}) to \textsc{tv} series, but as a mean
among other modalities to structure its contents.

In this paper, we propose to use the visual structure of narrative
movies to perform audiovisual speaker diarization within each
dialogue scene of \textsc{tv} series episodes.

Diarization on such movies presents some specific difficulties due to
audio-video asynchrony that limit the performance of video-only based
\textsc{sd} systems: the current speaker may not be filmed, the camera
focusing on the reaction of the character he is talking to. These
highly unpredictable asynchrony issues make necessary the joint use of
audio and video features.

On the other hand, movie dialogue scenes exhibit formal regularities
at a visual level, with two alternating and recurring shots, each one
corresponding to one of the two speakers involved. Once automatically
detected, such patterns could limit the interactivity scheme in which
diarization is performed.

This paper focus on speaker diarization in \textsc{tv} series. We
propose to perform independently audio and video-based speaker
diarization, before merging the resulting partitions of the spoken
segments in an optimal way. The two modalities are expected to be
uncorrelated in their respective mistakes, and to compensate each
other.

The way dialogue scenes are visually detected is described in
Section~\ref{sec:dialdetect}; the method used to perform mono-modal
speaker diarization is described in section~\ref{sec:monomod} and the
way the two resulting partitions of the utterance set are combined is
described in section~\ref{sec:multimod}. Experimental results are
given and discussed in section~\ref{sec:expres}.

\section{DIALOGUE SCENES VISUAL DETECTION}
\label{sec:dialdetect}

Relying on specific shot patterns, the detection of the dialogue
scenes requires the whole video stream be split into shots, before
these ones are compared and labelled according their similarities.

\subsection{Shot cut and shot similarity detection}
\label{ssec:segmsim}

The whole video stream can be regarded as a sequence of fixed images
(or frames) displayed on the screen at a constant rate able to
simulate for human eyes the continuity of motion. Moreover, a video
shot, as stated in~\cite{koprinska2001temporal}, is defined as an
``unbroken sequence of frames taken from one camera''. A new shot can
then be detected by comparing the current image to the next one: a
substantial difference between two temporally adjacent images is
indicative of a shot cut. Conversely, the current shot is considered
as similar to a past one if the first image of the former
substantially looks like the last one of the latter.

Both tasks, shot cut detection as well as shot similarity detection,
rely on image comparison. For this comparison purpose, images are
described by using 3-dimension histograms of the image pixel values in
the \textsc{hsv} color space. Comparison between images is then
performed by evaluating the correlation between the corresponding
color histograms. Nonetheless, different images may share the same
global color histogram, resulting in a irrelevant similarity:
information about the spatial distribution of the colors on the image
is then reintroduced by splitting the image into 30 blocks, each
associated to its own color histogram; block-based comparison between
image is then processed as described in~\cite{koprinska2001temporal}.

The two correlation thresholds needed to perform both tasks, shot cut
detection as well as shot similarity detection, are estimated by
experiments on a development set of~\textsc{tv} series episodes (see
Subsection~\ref{ssec:resshot}).

\subsection{Dialogue visual patterns extraction}
\label{sec:patterns}

Once shots are extracted and similar ones are detected, shot patterns
typical of short dialogue scenes can be detected. Let $\Sigma = \{l_1,
..., l_m\}$ be a set of possible shot labels, two shots sharing the
same label if they are hypothesized as similar.

The following regular expression $r(l_1, l_2)$ corresponds to a subset
of all the possible shot sequences $\Sigma^* = \bigcup_{n \geqslant 0}
\Sigma^n$:

\begin{equation}
  r(l_1, l_2) = \Sigma^* l_1(l_2l_1)^+ \Sigma^*
  \label{eq:patt}
\end{equation}

The set $\mathcal{L}(r(l_1, l_2)$ of sequences captured by the regular
expression~\ref{eq:patt} corresponds to shot label sequences
containing an occurrence of $l_2$ inserted between two occurrences of
$l_1$, with a possible repetition of the alternation $(l_2, l_1)$,
whatever be the previous and following shot labels. Such a regular
expression formalizes the ``two-alternating-and-recurring-shots''
pattern mentioned in Section~\ref{sec:intro} as typical of dialogue
scenes involving two characters.

\begin{figure}[htb]
  \vspace{0.5cm}
  \begin{minipage}[b]{1.0\linewidth}
    \centering
    \centerline{\includegraphics[width=9cm]{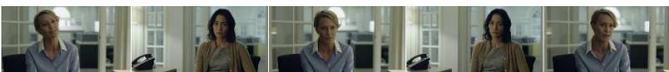}}
  \end{minipage}
  \caption{{\it Example of shot sequence $...l_1l_2l_1l_2l_1...$
      captured by the regular expression~\ref{eq:patt} for two shot
      labels $l_1$ and $l_2$.}}
  \vspace{0.5cm}
  \label{fig:patt1}
\end{figure}

Figure~\ref{fig:patt1} shows an example of a shot sequence matching the
regular expression~\ref{eq:patt}.

A movie can be described as a finite sequence $\mathbf{s} = s_1...s_k$
of $k$ shot labels, with $s_i \in \Sigma$. The set of patterns
$\mathcal{P}(\mathbf{s}) \subseteq \Sigma^2$ associated with the movie
shot sequence $\mathbf{s}$ can be defined as follows:

\begin{equation}
  \mathcal{P}(\mathbf{s}) = \{ (l_1, l_2) \in \Sigma^2 \ | \ \mathbf{s}
    \in r(l_1, l_2) \}
\end{equation}

The set of patterns $\mathcal{P}(\mathbf{s})$ contains all the pairs
of shots alternating with each other according to the
rule~\ref{eq:patt}.

In order to increase the speech coverage of such visual patterns,
isolated expressions of the alternating shot pairs involved in a
pattern are also taken into account.

The scenes of our corpus movies that match the regular
expression~\ref{eq:patt} appear to contain relatively few speech
(13.15 seconds in average) but cover more than half (53.10\%) of the
total amount of speech of a movie. Not surprisingly, the number of
speakers involved in such scenes with two alternating shots is close
to two speakers (1.84), with a standard deviation of 0.57: the scenes
with only one speaker are mainly the shortest ones, where the
probability that one of the two speakers remains silent increases.

\section{MONO-MODAL SPEAKER DIARIZATION}
\label{sec:monomod}

Widely available, the movie subtitles are here used to approximate
utterance boundaries. As an exact transcription of each utterance,
they usually match it temporally, despite some slight and
unpredictable latency before they are displayed and after they
disappear. When such a latency was too high, the utterance boundaries
were manually adjusted. Moreover, each subtitle is generally
associated with a single speaker, and in the remaining cases where two
speakers are involved in a single subtitle, speech turns are
indicated, allowing to split the whole subtitle into the two
corresponding utterances.

\subsection{Audio and visual features}
\label{ssec:features}

Once delimited, utterances can be described using either acoustic
or visual features.

The acoustic parameterization of utterances relies, as a state of the
art technique used in the speaker verification field, on the i-vectors
model (\cite{dehak2011front}). After 19 cepstral coefficients plus
energy are extracted, a 64-components \textsc{gmm/ubm} is trained on
the whole corpus (described in Subsection~\ref{ssec:corpus}); the
total variability matrix is then trained on all the spoken segments of
the currently processed movie, and 20-dimension i-vectors are finally
extracted, each associated with a single utterance. I-vectors are
extracted using the \textsc{alize} toolkit described
in~\cite{bonastre2005alize}.

On the other hand, the visual parameterization of an utterance relies
on its temporal distribution over the shots, as labelled according
their similarities (see subsection~\ref{ssec:segmsim}).

Considering the set $\Sigma = \{ l_1, ..., l_m \}$ of shot labels
involved in a movie, the $i-$th dimension of $\mathbb{R}_+^m $ is
associated to the $i-$th shot label. Each utterance $\mathbf{u} =
(u_0, ..., u_m)$ is then described as an $m-$dimension vector, where
the $i-$th component $u_i$ corresponds to the overlapping time in
seconds between the utterance $\mathbf{u}$ and the shot label $l_i$.

\begin{figure}[htb]
  \begin{minipage}[b]{1.0\linewidth}
    \centering
    \centerline{\includegraphics[width=7cm]{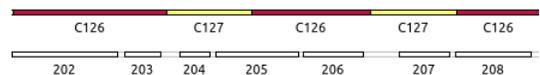}}
  \end{minipage}
  \caption{{\it Shot sequence
      $...c_{126}c_{127}c_{126}c_{127}c_{126}...$ for two shot labels
      $c_{126}$ and $c_{127}$ (top line) with the covered utterances
      (bottom line).}}
  \label{fig:patt2}
\end{figure}

Figure~\ref{fig:patt2} shows the distribution over time of the two
alternating shots of Figure~\ref{fig:patt1}, here labelled $(c_{126},
c_{127})$. The top line reports the alternation of both shots over
time; the bottom line contains the seven utterances covered by the
sequence.

For instance, The utterance $\mathbf{u^{(205)}}$ overlaps the two
shots and will then be set to 1.56 (seconds) for its $126-$th
component, to 1.16 for its $127-$th component, and to zero for all the
other ones.

\subsection{P-median clustering}
\label{ssec:pmed}

The $n$ utterances covered by a particular pattern can then be
described either according audio-only features, resulting in a set
$\mathcal{U}_a$ of $n$ 20-dimension i-vectors or by using visual-only
features, resulting in a set $\mathcal{U}_v$ of $n$ $m-$dimension
vectors, where $m$ denotes the number of shot labels in the movie.

Both sets $\mathcal{U}_a$ and $\mathcal{U}_v$ are first partitioned
into two clusters each, the average number of speakers by pattern
(1.84), as well as its standard variation, allowing such an \textit{a
  priori} assumption.

With such a fixed number of clusters, the partition problem can be
modelled using the $p-$median problem. The $p-$median problem
(\cite{hakimi1964optimum, hakimi1965optimum}) belongs to the family of
facility location problems: $p$ facilities must be located among
possible candidate sites such that the total distance between demand
nodes and the nearest facility is minimized.

The $p-$median problem can be transposed into the cluster analysis
context (\cite{klastorin1985p}) with a predefined number of
classes. The instances to cluster into $p$ classes correspond to the
demand nodes and each instance may be chosen as one of the $p$ class
centers. Choosing the centers so as to minimize the total distance
between the instances and their nearest center results in compact
classes with medoid centers.

Considering the set $\mathcal{U}$ of $n$ utterances covered by a
pattern, the clustering problem can be modelled using the following
binary decision variables: $x_i = 1$ if the $i-$th utterance $u^{(i)}$
is selected as one of the $p$ cluster centers, $x_i = 0$ otherwise;
$y_{ij} = 1$ if $u^{(i)}$ is assigned to the cluster center $u^{(j)}$,
$y_{ij} = 0$ otherwise. The model constants are the number of centers
$p$ as well as the distance coefficients $d_{ij}$ between the
utterances $u^{(i)}$ and $u^{(j)}$. The distance metric is the
euclidean one in the case of the video-based utterance vectors and the
normalized euclidean distance in the case of audio-based utterance
i-vectors.

The $p-$median clustering problem can then be modelled as the
following integer linear program, closely related to the program
described in~\cite{dupuy2012vectors, dupuy2014recent}:

\[
  \text{(P1)}
  \left \{
  \begin{aligned}
    & \min \left ( \sum_{i = 1}^n \sum_{j=1}^n d_{ij} y_{ij} \right ) \\
    \mathrm{s.t.} 
    & \left \bracevert
    \begin{aligned}
      & \sum_{j = 1}^n y_{ij} = 1 & i = 1, ..., n \\
      & \sum_{i = 1}^n x_i = p \\
      & y_{ij} \leqslant x_i & i = 1, ..., n; j = 1, ..., n \\
      & x_i \in \{0, 1\} & i = 1, ..., n \\
      & y_{ij} \in \{0, 1\} & i = 1, ..., n; j = 1, ..., n \\
    \end{aligned}
    \right. \\
  \end{aligned}
  \right.
\]

The first constraints $\sum_{j = 1}^n y_{ij} = 1$ ensures that each
utterance is assigned to exactly one center; the second one $\sum_{i =
  1}^n x_i = p$ that exactly $p$ centers are chosen; the third ones
$y_{ij} \leqslant x_i$ prevent an utterance from being assigned to a
non-center one.

Setting $p := 2$ and solving twice the integer linear program
$\text{(P1)}$, once for the utterance set $\mathcal{U}_a$ described by
audio features, and then for the utterance set $\mathcal{U}_v$ relying
on visual features, results in two distinct bipartitions of the same
utterance set.

\section{MULTI-MODAL COMBINATION}
\label{sec:multimod}

\subsection{Optimal matching fusion}
\label{ssec:optmatch}

The two bipartitions of the utterance set are then merged by solving
the classical maximum weighted matching in a bipartite graph.

\begin{figure}[htb]
  \begin{minipage}[b]{1.0\linewidth}
    \centering
    \vspace{0.25cm}
    \begin{pspicture}(8, 2.45)
      \psline[linewidth=.1]{->}(4.75, 1.225)(5.25, 1.225)
      \psframe[framearc=.2, linestyle = dotted](0.3, 0.1)(1.7, 2.45)
      \rput(1, 2){$\mathbf{u^{(1)}}$}
      \rput(0.75, 1.6){$\mathbf{u^{(2)}}$}
      \rput(1.35, 1.6){$\mathbf{u^{(3)}}$}
      \pscircle(1, 1.75){0.65}
      \rput(1, 0.5){$\mathbf{u^{(4)}}$}
      \pscircle(1, 0.5){0.35}
      \rput(0.2, 0){$\mathcal{Q}_a$}
      \rput(0, 1.75){\tiny $Q_a^{(1)}$}
      \rput(0, 0.5){\tiny $Q_a^{(2)}$}
      \psframe[framearc=.2, linestyle = dotted](2.45, 0.1)(3.55, 2.45)
      \rput(3, 2.15){$\mathbf{u^{(1)}}$}
      \rput(3, 1.75){$\mathbf{u^{(2)}}$}
      \pscircle(3, 1.9){0.5}
      \rput(3, 0.85){$\mathbf{u^{(3)}}$}
      \rput(3, 0.45){$\mathbf{u^{(4)}}$}
      \pscircle(3, 0.65){0.5}
      \rput(3.75, 0){$\mathcal{Q}_v$}
      \rput(3.9, 1.9){\tiny $Q_v^{(1)}$}
      \rput(3.9, 0.65){\tiny $Q_v^{(2)}$}
      \psline[linecolor=gray]{-}(1.65, 1.75)(2.5, 1.9)
      \psline[linecolor=gray]{-}(1.65, 1.75)(2.5, 0.65)
      \psline[linecolor=gray]{-}(1.35, 0.5)(2.5, 1.9)
      \psline[linecolor=gray]{-}(1.35, 0.5)(2.5, 0.65)
      \rput(2, 1.92){\tiny 2}
      \rput(2, 0.48){\tiny 1}
      \rput(1.5, 0.85){\tiny 0}
      \rput(2.35, 1){\tiny 1}
      \psframe[framearc=.2, linestyle = dotted](6.45, 0.1)(7.55, 2.45)
      \rput(7, 2.15){$\mathbf{u^{(1)}}$}
      \rput(7, 1.75){$\mathbf{u^{(2)}}$}
      \pscircle(7, 1.9){0.5}
      \rput(7, 0.5){$\mathbf{u^{(4)}}$}
      \pscircle(7, 0.5){0.35}
      \rput(6.25, 0){$\mathcal{Q}$}
      \rput(6.15, 1.9){\tiny $Q^{(1)}$}
      \rput(7.9, 0.5){\tiny $Q^{(2)}$}
    \end{pspicture}
  \end{minipage}
  \caption{{\it Set partitions fusion using maximum weighted matching in a bipartite graph}}
  \label{fig:comb}
\end{figure}
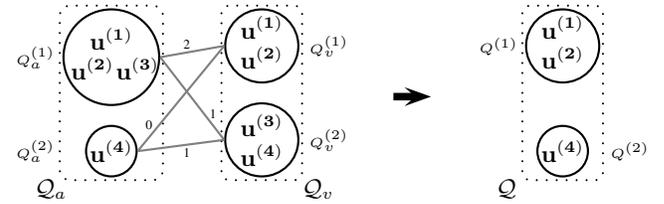

On Figure~\ref{fig:comb}, the set of utterances $\mathcal{U} = \{
\mathbf{u^{(1)}}, \mathbf{u^{(2)}}, \mathbf{u^{(3)}}, \mathbf{u^{(4)}}
\}$ is twice partitioned using the audio and video modalities,
resulting in two different partitions $\mathcal{Q}_a = \{ Q_a^{(1)},
Q_a^{(2)} \}$ and $\mathcal{Q}_v = \{ Q_v^{(1)}, Q_v^{(2)} \}$.

A bipartite weighted graph $\mathcal{G} = (\mathcal{Q}_a,
\mathcal{Q}_v, \mathcal{E})$, where $\mathcal{E} = \mathcal{Q}_a
\times \mathcal{Q}_v$, can then be defined by assigning to each edge
$(Q_a^{(i)}, Q_v^{(j)}) \in \mathcal{E}$ a weight $w_{ij}$
corresponding to the sum of the duration of the utterances that the
sets $Q_a^{(i)}$ and $Q_v^{(j)}$ have in common.

The edges of the bipartite graph on Figure~\ref{fig:comb} are thus
weighted by assuming a same duration of 1 for all the utterances
$\mathbf{u^{(1)}}, ..., \mathbf{u^{(4)}}$.

The best matching between both partitions consists in choosing
non-adjacent edges (without any node in common) so that the sum of
their weights is maximized. By using a decision variable $y_{ij}$ such
that $y_{ij} = 1$ if the edge $(Q_a^{(i)}, Q_v^{(j)})$ is chosen,
$y_{ij} = 0$ otherwise, the problem can be modelled as follows for two
bipartitions:

\[
  \text{(P2)}
  \left \{
  \begin{aligned}
    & \max \left ( \sum_{i = 1}^2 \sum_{j = 1}^2 w_{ij} y_{ij} \right ) \\
    \mathrm{s.t.} 
    & \left \bracevert
    \begin{aligned}
      & \sum_{j = 1}^2 y_{ij} \leqslant 1 & i = 1, 2 \\
      & \sum_{i = 1}^2 y_{ij} \leqslant 1 & j = 1, 2 \\
      & y_{ij} \in \{0, 1\} & (i, j) \in \{ 1, 2 \}^2 \\
    \end{aligned}
    \right. \\
  \end{aligned}
  \right.
\]

The first and second constraints ensure that only non adjacent edges
will possibly be chosen.

In the example of Figure~\ref{fig:comb}, the best choice consists
in assigning $Q_a^{(1)}$ to $Q_v^{(1)}$ and $Q_a^{(2)}$ to
$Q_v^{(2)}$, for a total cost of 3.

Once the matching choice is made by solving the problem $\text{(P2)}$,
the matching subsets are intersected, resulting in a new set
$\mathcal{Q}$ of subsets of $\mathcal{U}$ corresponding to cases of
agreement between the two modalities: the subsets obtained are
expected to contain segments both acoustically close to each other and
uttered as the corresponding speaker is filmed. Conversely, the
residual segments ($\mathbf{u^{(3)}}$ in the example of
Figure~\ref{fig:comb}) are discarded as cases of disagreement between
the audio and visual modalities, either because the utterance is
acoustically atypical or because of asynchrony between the utterance
and the character currently filmed.

\subsection{Reallocation of discarded utterances}
\label{ssec:realloc}

The residual utterances are finally reallocated to the closest medoids
of the refined clusters resulting from the combination of the audio
and visual modalities.

This stage of reallocation relies on the audio-only features of the
remaining utterances: possibly discarded because of their visual
asynchrony, such utterances might not be correctly reallocated by
relying on the visual modality. On the other hand, using the audio
modality to achieve such a reallocation is expected to be more robust
than when performing the audio-only clustering described in
subsection~\ref{ssec:pmed}: by using medoids of clusters refined by
the use of the video modality, some errors made during the audio-only
stage are expected to be here avoided. Moreover, medoid, being less
sensitive to outliers than centroid, is expected to properly handle
the case of impure clusters containing isolated misclassified
utterances resulting from a joint error of both modalities.

\section{EXPERIMENTS AND RESULTS}
\label{sec:expres}

\subsection{Corpus}
\label{ssec:corpus}

For experimental purpose, we used the first seasons of three
\textsc{tv} series: \textit{Breaking Bad} (abbreviated \textit{bb}),
\textit{Game of Thrones} (\textit{got}), and \textit{House of Cards}
(\textit{hoc}). We manually annotated three episodes of each series by
indicating shot cuts, similar shots, speech segments as well as the
corresponding speakers. The total amount of speech in these nine
episodes represents a bit more than three hours (3:12).

\subsection{Shot cuts and shot similarities detection}
\label{ssec:resshot}

The evaluation of shot cut detection relies on a classical F1-score
(\cite{boreczky1996comparison}). For the shot similarity detection
task, an analogous F1-score is used: for each shot, the list of shots
hypothesized as similar to the current one is compared to the
reference list of similar shots; if both lists intersect in a
non-empty set, the shot is considered as correctly paired with its
list. As both these image processing tasks require thresholds
estimation, a development subset of six episodes is here used. Average
results on \textsc{dev} and \textsc{test} sets are reported in
Table~\ref{resshot}.

\begin{table}[h]
  \caption{\label{resshot}{\it Average results obtained for shot cut and shot similarity detection}}
  \vspace{2mm}
  \centering
  \begin{tabular}{|c|c|ccc|}
    \hline
    & \textbf{shot cut} & \multicolumn{3}{|c|}{\textbf{shot similarity}} \\
    \cline{2-5}
    & F1-score & precision & recall & F1-score \\
    \hline
    avg. \textsc{dev} & \textbf{0.97} & 0.90 & 0.88 & \textbf{0.89} \\
    \hline
    \hline
    avg. \textsc{test} & \textbf{0.99} & 0.91 & 0.90 & \textbf{0.90} \\
    \hline
  \end{tabular}
\end{table}

The results obtained for the shot similarity detection task (F1-score
amounting to 0.90) are expected to make reliable the visual detection
of alternating, recurring shots as typical of short dialogue scenes
involving two characters.

\subsection{Speaker diarization within dialogue scenes}
\label{ssec:spkdiar}

Speaker diarization, performed within each dialogue scene as
hypothesized from visual clues, is evaluated using the single show
\textsc{der} (\cite{rouvier2013open}): the \textsc{der} is computed
for each dialogue scene before the results are averaged according each
dialogue duration. Results are given for the mono-modal speaker
diarization step for both modalities, audio and video. The optimal
matching (denoted \textit{om}) performed during the multi-modal fusion
step is evaluated in two ways: first by discarding from scoring the
utterances for which the two modalities disagree (denoted
\textit{om-ra}). In this case, the resulting speech coverage of the
scored utterances in indicated in percentage in parenthesis. Moreover,
results are also given when the optimal matching between both
modalities is followed by a step of audio reallocation of the
remaining utterances (denoted \textit{om+ra}). For the sake of
comparison, the results obtained by optimizing jointly in a weighted
sum (denoted \textit{ws}) the two $p-$median mono-modal objective
functions are also reported. Finally, an oracle score is estimated by
labelling the utterances according the reference speaker when at least
one of both modalities succeeds in retrieving it.

\begin{table}[h]
  \caption{\label{resspkdiar}{\it Single show Diarization Error Rate
      obtained for all episodes}}
  \vspace{2mm}
  \centering
  \begin{tabular}{|c|cc||c||ccc|}
    \hline
    & \multicolumn{2}{|c||}{\textbf{mono-modal}} &
    \textbf{oracle} & \multicolumn{3}{|c|}{\textbf{multi-modal}}\\
    \cline{2-7}
    & \textit{audio} & \textit{video} & & \textit{om-ra} &
    \textit{om+ra} & \textit{ws} \\
    \hline
    \hline
    \textit{bb-1} & 25.2 & 26.9 & 8.3 & 18.0 {\tiny \textbf{(67.7)}} &
    24.0 & 26.9 \\
    \hline
    \textit{bb-2} & 26.6 & 24.5 & 8.2 & 17.2 {\tiny \textbf{(69.7)}} &
    20.4 & 24.5 \\
    \hline
    \textit{bb-3} & 26.8 & 26.9 & 9.6 & 17.1 {\tiny \textbf{(67.4)}} &
    24.7 & 27.3 \\
    \hline
    \textit{got-1} & 22.6 & 24.7 & 7.6 & 13.1 {\tiny \textbf{(69.2)}} &
    21.1 & 24.5 \\
    \hline
    \textit{got-2} & 28.7 & 27.7 & 10.2 & 20.0 {\tiny \textbf{(68.2)}} &
    25.9 & 27.0 \\
    \hline
    \textit{got-3} & 12.8 & 29.4 & 5.3 & 9.9 {\tiny \textbf{(71.1)}} &
    13.1 & 28.2 \\
    \hline
    \textit{hoc-1} & 17.5 & 21.9 & 3.8 & 10.0 {\tiny \textbf{(71.6)}} &
    17.7 & 22.2 \\
    \hline
    \textit{hoc-2} & 21.4 & 29.4 & 10.2 & 15.4 {\tiny \textbf{(70.6)}} &
    20.8 & 29.4 \\
    \hline
    \textit{hoc-3} & 20.6 & 25.6 & 6.9 & 12.8 {\tiny \textbf{(70.2)}} &
    20.6 & 25.4 \\
    \hline
    \hline
    avg. & \textbf{22.5} & \textbf{26.3} & \textbf{7.8} & \textbf{14.8} {\tiny \textbf{(69.5)}} & \textbf{20.9} & 26.2 \\
    \hline
  \end{tabular}
\end{table}

As can be seen, the results obtained by performing mono-modal speaker
diarization are in average slightly better for the audio modality than
for the video one. Nonetheless, the computed oracle shows that both
modalities are not redundant: by managing to combine them perfectly,
the \textsc{der} would decrease dramatically (from 22.5\% to 7.8\% for
the audio modality, and from 26.3\% to 7.8\% for the video one), which
confirms that both these modalities are highly complementary for the
speaker diarization task and that the errors made are not correlated.

Moreover, when both modalities are combined, resulting in a new
partial clustering of the utterance set, the \textsc{der} remains
relatively low if about 30\% of the utterances, corresponding of cases
of disagreement between both modalities, are discarded from the
evaluation (\textsc{der} amounting to 14.8\% for 69.5\% of speech
covered).

Not surprisingly, while processing the critical 30\% remaining
utterances, the \textsc{der} tends to increase (from 14.8\% to 20.9\%)
but is still lower than the \textsc{der} obtained for the single audio
modality (22.5\%), a relative improvement of 7.11\%.

\section{CONCLUSION}
\label{ssec:conclusion}

In this paper, we proposed to perform audiovisual speaker diarization
within short scenes of \textsc{tv} series visually hypothesized as
dialogues between two characters. Speaker diarization is first
performed separately for audio and visual features of the utterances
by using the $p-$median model, before both resulting bipartitions
of the utterance set are optimally matched in new clusters
corresponding to cases of agreement between both modalities. The
isolated remaining utterances for which both modalities disagree are
then acoustically assigned to the closest centroids of the newly
created clusters, expected to be more robust than when based on an
audio-only approach. The experimental results obtained by using both
modalities turn out to outperform those obtained by purely mono-modal
approaches.

\vfill\pagebreak

\bibliographystyle{IEEEbib}
\bibliography{refs}

\end{document}